\documentclass[12pt]{article}
\usepackage[dvips]{graphicx}
\def\thefootnote{\fnsymbol{footnote}}

\newcommand{\eq}{\begin{equation}}
\newcommand{\en}{\end{equation}}
\newcommand{\eqa}{\begin{eqnarray}}
\newcommand{\ena}{\end{eqnarray}}

\newcommand{\mc}{\multicolumn}

\begin{document}
\begin{titlepage}
\vskip0.5cm
\begin{flushright}
HUB-EP-99/18\\
\end{flushright}
\vskip0.5cm
\begin{center}
{\Large\bf High precision Monte Carlo study} 
\vskip0.3cm
{\Large\bf of the 3D XY-universality class}
\end{center}
\vskip 1.3cm
\centerline{
 M. Hasenbusch \footnote{e--mail: hasenbus@physik.hu-berlin.de} 
 and T. T\"or\"ok \footnote{e--mail: toeroek@physik.hu-berlin.de}}
 \vskip 1.0cm
 \centerline{\sl Humboldt Universit\"at zu Berlin, Institut f\"ur Physik}
 \centerline{\sl Invalidenstr. 110, D-10115 Berlin, Germany}
 \vskip 1.cm

\begin{abstract}
We present a Monte Carlo study of the two-component
$\phi^4$ model on the simple cubic lattice in 
three dimensions.  By suitable tuning of the coupling 
constant $\lambda$ we eliminate leading order corrections to scaling.
High statistics simulations using finite size scaling techniques 
yield $\nu=0.6723(3)[8]$ and $\eta=0.0381(2)[2]$, where the statistical
and systematical errors are given in the first and second bracket,
respectively.
These results are more precise 
than any previous theoretical estimate of the critical exponents
for the 3D XY universality class.
\vskip0.2cm
\end{abstract}
\end{titlepage}

\setcounter{footnote}{0}
\def\thefootnote{\arabic{footnote}}

\section{Introduction}
The 3D XY universality class is unique in the respect that 
experimental estimates for critical exponents are
more precise than any theoretical estimate.
These experiments are performed in the neighbourhood of
the super-fluid transition 
of $^4$He. The specific heat  or the super-fluid
density is measured as a function of the temperature 
\cite{helium1,helium2,helium3}.

In the present study we try to close the gap between theory and 
experiment by a high statistics Monte Carlo simulation 
of the two-component
$\phi^4$ (or Landau-Ginzburg) model
on  a three dimensional simple cubic lattice.
The action is given by
\eq
S = \sum_x \{- 2 \kappa \; \sum_{\mu} \vec{\phi}_x \vec{\phi}_{x+\hat \mu}
+\vec{\phi}_x^2
+ \lambda \; (\vec{\phi}_x^2 -1)^2 \} \;\; ,
\label{action}
\en
where
the field variable $\vec{\phi}_x$
is a vector with two real components and $x=(x_1,x_2,x_3)$, 
where $x_i$ is integer, labels the lattice
sites. $\mu$ labels the directions and $\hat\mu$ is a unit-vector in
$\mu$-direction. The Boltzmann factor is $\exp(-S)$.
For $\lambda=0$ we get the Gaussian model on the lattice. In the
limit $\lambda=\infty$ the XY-model is recovered.

In addition to statistical errors Monte Carlo estimates of 
critical exponents are affected 
by systematical errors that result from corrections to scaling. 
These systematical errors can be reduced 
(in a finite size scaling study) by   
increasing the linear size $L$ of
the lattices that are simulated. 
A more elegant approach is to remove
corrections by a suitable choice of the action.
Recently it was demonstrated that leading order corrections to scaling 
can be removed by a suitable tuning of the coupling constant $\lambda$
in the 
one-component $\phi^4$ theory on the lattice \cite{spain,us,MH}.
Leading order corrections to scaling are proportional to $\xi^{-\omega}$
($L^{-\omega}$ in finite size scaling),
where $\xi$ is the correlation length and $\omega\approx 0.8 $.

The paper is organised as follows: In section 2 we discuss the observables that are measured. In section 3 we explain the algorithm that has 
been used for the simulation and we summarise the simulation parameters.
In section 4 the data are analysed. In section 5
our results for exponents 
are compared with experimental and theoretical estimates given in the 
literature. In section 6 we give our conclusions and an outlook.

\section{The measured quantities}
In the case of the one-component model the Binder cumulant turned out 
to be a good indicator for corrections to scaling \cite{MH}. 
The Binder cumulant is defined by
\begin{equation}
 U = \frac{<(\vec{m}^2)^2>}{<\vec{m}^2>^2} \;\; ,
\end{equation}
where 
\begin{equation}
 \vec{m} = \frac{1}{V} \sum_x \vec{\phi}_x \;\; 
\end{equation}
is the  magnetisation per lattice site 
of a given configuration. The volume is  
$V=L^3$. 
In the following we will always consider systems with periodic 
boundary conditions.
In ref. \cite{MH} the Binder cumulant was computed at a fixed value 
of the ratio of partition functions $Z_a/Z_p$. $Z_a$ is the partition 
function for anti-periodic boundary conditions and $Z_p$ for periodic
boundary conditions. This ratio can also be computed for an arbitrary 
number of components. For a simulation of the XY model see ref. 
\cite{GoHa}. However in the present paper we have replaced $Z_a/Z_p$
by the dimension-less ratio $\xi_{2nd}/L$ 
because the second moment correlation length $\xi_{2nd}$
is easier to implement as $Z_a/Z_p$. Note that $\xi/L$, where $\xi$
is the exponential correlation length on a strip of width $L$, 
was used in the pioneering work of Nightingale \cite{Night} on the 
phenomenological renormalization 
group approach.

The second moment correlation length is defined by
\begin{equation}
 \xi_{2nd} = \left(\frac{\chi/F -1}{4 \sin^2(\pi/L)} \right)^{1/2} \;\;,
\end{equation}
where the 
magnetic susceptibility is given by
\begin{equation}
\chi= {V} \; \langle \vec{m}^2 \rangle
\end{equation}
and 
\begin{equation}
 F = \frac{1}{V} \;
 \langle|\sum_x \exp\left(\;i \;
 \frac{ 2 \pi x_1}{L} \right) \; \vec{\phi}_x|^2
 \rangle
\end{equation}
is the Fourier transform of the correlation function at minimal momentum.
In the simulation we averaged over all three directions to reduce the 
statistical error. Note that in the following $\xi_{2nd}$ is always 
evaluated at a finite value of $L$ and not for the thermodynamic limit.

We performed some simulations of the one-component model to compare 
$\xi_{2nd}/L$ and $Z_a/Z_p$. 
We found that the physical as well as statistical properties of 
$\xi_{2nd}/L$ and $Z_a/Z_p$ are similar. 

In order to compute observables in the neighbourhood of the simulation
parameter $\kappa_s$ we computed the first two coefficients of the Taylor
expansion in $\kappa-\kappa_s$. We always checked that the errors made 
by the truncation of the Taylor series are much smaller than the 
statistical errors of the quantities that were computed.

\section{The Simulations}
\subsection{The Monte Carlo algorithm}
We generalise the idea of Brower and Tamayo \cite{BrTa} to simulate 
the one-component $\phi^4$ theory. They use the Swendsen-Wang cluster 
algorithm \cite{SwWa}
to update the sign of the field $\phi$. In order to obtain an 
ergodic update they supplement the cluster-update with a Metropolis 
update that also allows to update the modulus of the field.
In our case we use the single cluster algorithm \cite{Wolff} only to 
update the direction of the field. The modulus is updated with the
Metropolis algorithm.
Let us briefly recall the steps of the single cluster algorithm applied
to the two-component $\phi^4$ theory.
First a direction  $\vec{n}$ is chosen
\begin{equation}
n_1 = \sin(2\pi \theta) \;\;\; , \;\; n_2 = \cos(2\pi \theta) \;\; ,
\end{equation}
where $\theta$ is a random-number that is uniformly distributed in
$[0,1)$. Next randomly a site of the lattice  is picked as seed of the 
cluster.  The cluster is build recursively. New sites enter the cluster
when they freeze onto their neighbours that are already members of
the cluster. 
The freezing probability is $p_f=1-p_d$ with
\begin{equation}
p_d =
\mbox{min}\left[1,\; \exp(- 4 \kappa \;
(\vec{n} \vec{\phi}_x) \; (\vec{n} \vec{\phi}_y) ) \right] \;\; .
\end{equation}
The fields of all sites in the cluster are reflected
\begin{equation}
\vec{\phi}_x' =  \vec{\phi}_x - 2 \; (\vec{n} \vec{\phi}_x) \; 
\vec{n} \;\;.
\end{equation}
The modulus of $\vec{\phi}$ is changed with a local Metropolis update.
A proposal for the field is generated by
\begin{equation}
\phi_{i,x}' = \phi_{i,x} + s (r_i - 0.5)
\end{equation}
for $i=1,2$,
where $r_i$ is a random-number that is uniformly distributed in $[0,1)$.
The acceptance probability is given by
\begin{equation}
A = \mbox{min}[1,\; \exp(S-S')] \;\;,
\end{equation}
where $S$ and $S'$ are the action for the original field and the proposal,
respectively.
We found that a step-size $s = 2$ yields an acceptance rate of about
$50 \%$. In one sweep we go trough the lattice in lexicographic order.

\subsection{The simulation parameters}
We performed simulations at a large range of $\lambda$ values and 
linear lattice sizes $L$. In table \ref{statist}
we give an overview of the simulation parameters
and the number of measurements for each set of simulation parameters.
Most of our simulations were performed on 200MHz Pentium Pro PCs 
running  under
Linux. The program is written in C. As random number generator 
we used our own implementation of G05CAF of the NAG-library. 
The total amount of CPU-time used for the simulations was about 3 years
on the 200MHz Pentium Pro PCs.

\begin{table}[h]
\caption[bx1]
{\label{statist}
\sl Summary of simulation parameters. In the first row we 
give the value of $\lambda$, in the second row the linear lattice size $L$ and 
in the third row the number of measurements divided by $3 \times 10^{6}$. 
}
\vskip 0.2cm
\begin{center}
\begin{tabular}{|l|l|l|}
\hline
$\lambda$ &  $L$                     & stat/$3  \times 10^{6} $   \\
\hline
   0.5    & 8,16                     & 17,5 \\
   1.0    & 6,8,10,12,14,16,18,20,22,24 & 50,10,10,10,10,10,10,10,10,11 \\
   1.5    & 8,16                     &13,6  \\
   1.7    & 8,12,24                  & 15,10,2.5 \\
   1.8    & 3,4,5,6,7,8,9,10,12,16   &20,67,67,20,40,15,45,30,15,8\\
   1.9    & 3,4,5,6,7,8,12,16,20,24  &33,27,20,20,15,20,10,10,11,10\\
   1.98   & 8,12,16,20,24            &20,15,10,11,15\\
   2.0    &3,4,5,6,7,8,9,10,11,12,13,14&
	  133,67,67,50,25,20,24,20,20,20,20,30\\
  &15,16,18,20,22,24,26,28,32,40,48& 20,20,25,25,25,20,16,15,22,10,10 \\
   2.2    & 3,4,5,6,7,8,9,10,12,16,24& 133,67,67,50,40,15,45,30,15,8,5 \\
   4.0    & 6,7,8,9,10,11,12,14,16,18,20,22,24 & 
	     50,20,10,10,10,9,10,10,10,10,10,10,10\\
\hline
\end{tabular}
\end{center}
\end{table}

Per measurement we performed  one sweep with the Metropolis algorithm 
and $m$ single cluster updates. The number of cluster updates was chosen 
roughly proportional to the linear lattice size $L$.  For some lattice 
sizes we searched for the $m$ that gives the optimal performance of the 
algorithm. For $L=48$ we found $m=40$ as optimal. 

\section{Analysing the data}
\subsection{The Binder cumulant and corrections to scaling}
We analysed the Binder cumulant at $\xi_{2nd}/L=0.5927$ fixed. This means 
that first (at fixed $\lambda$)  $\kappa_f$ is computed for that 
$\xi_{2nd}/L=0.5927$. Then the Binder cumulant is computed at 
$\kappa_f$. In the following we denote the Binder cumulant at 
$\xi_{2nd}/L=0.5927$ by $\bar{U}$.  
From preliminary simulations we know that $\xi_{2nd}/L=0.5927$ is a 
good approximation of
\begin{equation}
\xi_{2nd}/L^* = \lim_{L \rightarrow \infty} \;\; \xi_{2nd}/L|_{\kappa_c}
\;\;\; .
\end{equation}
The advantage of this approach is 
that we need not to search for $\kappa_c$ and that due to 
cross-correlations the statistical error of the Binder cumulant at 
$\xi_{2nd}/L=0.5927$ fixed is smaller than at a given value of $\kappa$.
(See e.g. ref. \cite{mc2}.)

For large $L$  $\bar{U}$ approaches a universal constant $\bar{U}^*$.
Leading order corrections are given by
\begin{equation}
\label{simple1}
 \bar{U}(L,\lambda) = \bar{U}^* + c_1(\lambda) \;\; L^{-\omega} \;\;.
\end{equation}
We fitted the data for all values of $\lambda$ simultaneously with 
this ansatz. The free parameters of this fit are 
$\bar{U}^*$, $\omega$ and $c_1(\lambda)$ for each value of $\lambda$.

The results for various minimal lattice sizes $L_{min}$ that have been 
included in the fit are 
summarised in table \ref{omega1}.
\begin{table}[h]
\caption[bx2]
{\label{omega1} \sl
Fit results for the Binder cumulant evaluated at $\xi_{2nd}/L=0.5927$
fixed. The ansatz is given in eq. (\ref{simple1}).
We  give results for various minimal lattice sizes $L_{min}$,
 $\omega$ is the correction to scaling exponent.}
\vskip 0.2cm
\begin{center}
\begin{tabular}{|c|c|c|c|}
\hline
\rule[0mm]{0mm}{4mm}
$L_{min}$ & $\chi^2$/d.o.f. &  $\bar{U}^*$  &  $\omega$ \\
\hline
 \phantom{0}6 &   5.42  &  1.24357(3)    &  0.786(6)\phantom{0} \\
 \phantom{0}8 &   2.34  &  1.24324(4)    &  0.775(6)\phantom{0} \\   
  10      &   2.15  &  1.24311(5)    &  0.788(10) \\
  12      &   1.80  &  1.24297(6)    &  0.782(14) \\
  14      &   1.75  &  1.24279(8)    &  0.790(20) \\
  16      &   1.86  &  1.24274(9)    &  0.819(31) \\
\hline
\end{tabular}
\end{center}
\end{table}
The values for $\chi^2$/d.o.f. stay rather large as $L_{min}$ 
is increased. We could not pin-point a particular problem that caused
this effect. On the other hand the result for the exponent $\omega$ is 
quite stable as $L_{min}$ is varied. 
As our final result for the correction to scaling exponent we quote 
$\omega=0.79(2)$. It is hard to give reliable estimates for the systematical
errors. At least the fact that the result for $\omega$ stays almost 
constant starting from  $L_{min}=6$ indicates that these errors should
be small. 

For $L_{min}=12$, $14$ and $16$ we
give the results for $c_1(\lambda)$ in table \ref{constant1}.
Linear interpolation of the result for $c_1$ at $\lambda=2.0$ and 
$\lambda=2.2$ yields $\lambda_{opt} = 2.046(9)$, $2.086(9)$ 
and $2.101(10)$ 
for $L_{min}=12$, $14$ and $16$, respectively. Where $\lambda_{opt}$ is 
defined by $c(\lambda_{opt})=0$. 
There is still an increase in
$\lambda_{opt}$ visible as $L_{min}$ increases.
We quote $\lambda_{opt}=2.10(1)[5]$
as our final result. As a rough estimate of systematical errors we give
(in the square brackets)
the 
difference of the result for  $L_{min}=12$ and $L_{min}=16$. 
\begin{table}[h]
\caption[bx3]
{
\label{constant1} \sl
The correction to scaling amplitude $c_1(\lambda)$ as a function
of $\lambda$ from fits with the ansatz (\ref{simple1}).
 We give the results for
three values of $L_{min}=12$, $14$ and $16$. 
}
\vskip 0.2cm
\begin{center}
\begin{tabular}{|c|c|c|c|}
\hline
 $\lambda$ & $L_{min}=12$   &  $L_{min}=14$   & $L_{min}=16$  \\
\hline
0.5\phantom{0}&\phantom{--}0.2152(83)&
	       \phantom{--0}0.2220(122) &
	       \phantom{--}0.2408(207) \\
1.0\phantom{0}&\phantom{--}0.0956(37)& 
	       \phantom{--}0.0999(59)&
	       \phantom{--}0.1094(101)\\   
1.5\phantom{0}&\phantom{--}0.0398(21)&
	       \phantom{--}0.0424(28)&
	       \phantom{--}0.0464(43)\phantom{0}\\
1.7\phantom{0}&\phantom{--}0.0229(12) &
	       \phantom{--}0.0280(32) &
               \phantom{--}0.0314(42)\phantom{0} \\
1.8\phantom{0}&\phantom{--}0.0153(8)\phantom{0}  &  
	       \phantom{--}0.0186(17)    & 
	       \phantom{--}0.0207(23)\phantom{0}  \\
1.9\phantom{0}&\phantom{--}0.0077(8)\phantom{0}  &   
	       \phantom{--}0.0099(11)    &  
	       \phantom{--}0.0114(15)\phantom{0}  \\
1.98          &\phantom{--}0.0038(7)\phantom{0}  &   
	       \phantom{--}0.0067(11)    &  
	       \phantom{--}0.0079(14)\phantom{0}  \\
2.0\phantom{0}&\phantom{--}0.0022(6)\phantom{0}  &   
	       \phantom{--}0.0043(9)\phantom{0}    &  
	       \phantom{--}0.0057(13)\phantom{0}  \\
2.2\phantom{0}& --0.0074(7)\phantom{0}          & 
		--0.0057(13)    &
		--0.0056(16)\phantom{0}  \\
4.0\phantom{0}& --0.0604(24)    & 
		--0.0601(37)    &
		--0.0649(63)\phantom{0}  \\
\hline
\end{tabular}
\end{center}
\end{table}

Following ref. \cite{MH} we tried to fit our data with the extended ansatz
\begin{equation}
\label{simple2}
 \bar{U}(L,\lambda) = \bar{U}^* + c_1(\lambda) \;\; L^{-\omega}
		     + c_2 \; c_1(\lambda)^2  \;\; L^{-2 \omega} \;\;.
\end{equation}
However it turned out that we have too few data with a large
difference $\bar{U}-\bar{U}^*$ to resolve $c_2$. 

Finally we fitted the difference of the Binder cumulant at $\lambda=2.0$
and $\lambda=2.2$ with the ansatz
\begin{equation}
\label{delta}
\bar{U}(L,\lambda=2.0) -\bar{U}(L,\lambda=2.2) = c \;\; L^{-\omega} \;\;.
\end{equation}
Results are given in the table \ref{constant}.  It turns out that the
$\chi^2$/d.o.f.  becomes order $1$ already for the very small $L_{min}=3$.
Also the value obtained for $\omega$ with this small $L_{min}$ is 
consistent with the result obtained above. Hence corrections beyond 
$L^{-\omega}$ do depend very little on $\lambda$ and are cancelled in
$\bar{U}(L,\lambda=2.0) -\bar{U}(L,\lambda=2.2)$. The same observation 
holds in the case of the one-component model \cite{MH}. 
\begin{table}[h]
\caption[bx4]
{
\label{constant} \sl
Fitting the difference of $\bar{U}$ at $\lambda=2.0$ and $\lambda=2.2$
with the ansatz (\ref{delta}).
}
\vskip 0.2cm
\begin{center}
\begin{tabular}{|c|c|c|c|}
\hline
$L_{min}$  &  $\omega$  &  $c_1(2.0)-c_2(2.2)$ &  $\chi^2$/d.o.f. \\
\hline
 3         & 0.787(18)  &   0.0106(3)    &  0.98       \\
 4         & 0.780(31)  &   0.0104(5)    &  1.09       \\
 5         & 0.794(43)  &   0.0107(9)    &  1.21       \\
\hline
\end{tabular}
\end{center}
\end{table}

\subsection{The critical line $\kappa_c(\lambda)$}

As approximation of the critical $\kappa_c$   we take  
$\kappa_f$ where
$\xi_{2nd}/L=0.5927$.  In table \ref{kritisch} we give the result for 
the largest lattice size available for each value of $\lambda$ that has 
been studied. Leading corrections are given by
\begin{equation}
 \kappa_f - \kappa_c = a \; L^{-1/\nu} + b \; L^{-1/\nu-\omega} \; + \;...
 \;\;\; .
\end{equation}
The constant $a$ should be very small since we have chosen 
$\xi_{2nd}/L=0.5927$ as a good approximation of $\xi_{2nd}/L^*$.
The value of $b$ depends on $\lambda$ and vanishes at $\lambda_{opt}$.
Nevertheless we assume pessimistically that errors decay with $L^{-1/\nu}$.
Systematical errors are then computed by comparing $\kappa_f$  at $L$ with 
$\kappa_f$ at $L/2$.  These errors are given in square brackets. Whenever
statistical errors reach a similar size as the systematical ones they are
quoted in addition in round brackets.
\begin{table}[h]
\caption[bx5]
{
\label{kritisch} \sl
Estimates of the critical $\kappa_c$ for all values of $\lambda$
that have been simulated. The value for $\lambda=\infty$ has been 
taken from ref. \cite{mc2}. The systematical errors are given in 
square brackets and 
statistical errors in round brackets. 
}
\vskip 0.2cm
\begin{center}
\begin{tabular}{|l|c|l|} \hline
\multicolumn{1}{|c|}{$\lambda$}
 & $L$ &
\multicolumn{1}{|c|}{2 $\kappa_{c}$} \\ \hline
0         &     & 0.33...   \\
0.5       & 16  & 0.4828[6]  \\
1.0       & 24  & 0.50754[7] \\
1.5       & 16  & 0.51197[7]  \\
1.7       & 24  & 0.51160[2] \\
1.8       & 16  & 0.51115[2] \\
1.9       & 24  & 0.510576(2)[7] \\
1.98      & 24  & 0.510049(1)[7] \\
2.0       & 48  & 0.5099049(6)[9]\\
2.2       & 24  & 0.508344(2)[4] \\
4.0       & 24  & 0.49243[5]  \\
$\infty$  &     & 0.454165(4) \\
\hline
\end{tabular}
\end{center}
\end{table}

\subsection{The exponent $\eta$}
We computed the exponent $\eta$ from the finite size behaviour 
of the magnetic susceptibility $\chi$ at either $\xi_{2nd}/L = 0.5927$ 
or $U=1.243$ fixed.  We denote the magnetic susceptibility at
$\xi_{2nd}/L$ or  $U$ fixed by $\bar{\chi}$. 
It scales as
\begin{equation}
\label{chifit1}
\bar{\chi} =  d  \,  L^{2-\eta} \;\; . 
\end{equation}
First we analysed our data for $\lambda=2.0$ which is close to 
$\lambda_{opt}$ and where we have accumulated most data.  Results
for $\xi_{2nd}/L$ fixed are given in table \ref{chixi1} and for 
$U$ fixed in table \ref{chiu1}. 

In both cases rather large $L_{min}$ are needed to reach a $\chi^2/$d.o.f.
close to 1. Since $\bar{\chi}$ at fixed $\xi_{2nd}/L$ has a smaller 
statistical error than  $\bar{\chi}$ at fixed $U$
also the statistical error of 
$\eta$ is smaller for $\xi_{2nd}/L$ fixed  than for $U$ fixed.

\begin{table}[h]
\caption[bx6]
{\label{chixi1} \sl Fits of the magnetic susceptibility at
$\xi_{2nd}/L=0.5927$ fixed with the ansatz (\ref{chifit1}).
}
\vskip 0.2cm
\begin{center}
\begin{tabular} {|c|c|c|c|}  \hline
$L_{min}$  & $d$ & $\eta$ & $\chi^{2}/$d.o.f. \\ \hline
 14 & 1.25629(20) & 0.03667(5)\phantom{0} &  10.28 \\
 24 & 1.25957(50) & 0.03742(11) &  \phantom{0}2.91 \\
 26 & 1.26067(61) & 0.03766(14) &  \phantom{0}0.70 \\
 28 & 1.26117(75) & 0.03777(17) &  \phantom{0}0.40 \\
\hline
\end{tabular}
\end{center}
\end{table}

\begin{table}[h]
\caption[bb27]
{\label{chiu1} \sl Fits of the magnetic susceptibility at
$U=1.243$ fixed with the ansatz (\ref{chifit1}).
}
\vskip 0.2cm
\begin{center}
\begin{tabular} {|c|c|c|c|} \hline
$L_{min}$  & $d$ & $\eta$ & $\chi^2/$ d.o.f. \\ \hline
 14 & 1.2598(6)\phantom{0}  & 0.03740(16) &   2.27 \\
 22 & 1.2628(13) & 0.03811(31) &   1.20 \\
 24 & 1.2644(16) & 0.03845(38) &   0.83 \\
 26 & 1.2625(20) & 0.03804(46) &   0.29 \\
\hline
\end{tabular}
\end{center}
\end{table}

Because we had to go to large $L_{min}$ with the simple ansatz 
(\ref{chifit1}) we added an analytic correction 
\begin{equation}
\label{chifit2}
\bar{\chi} =c +  d  \,  L^{2-\eta} \;\;\; .
\end{equation}
Note that also corrections  that
decay like $L^{-x}$ with $x\approx 2$ 
are effectively parametrised by this ansatz. Results for fits with this 
ansatz are given in table \ref{chixi2} for $\xi_{2nd}/L$ fixed and for
$U$ fixed in table \ref{chiu2}.  We see that a small $\chi^2/$d.o.f.
is already reached for  $L_{min}=7$ and $L_{min}=6$ respectively.
Despite the fact that  $\chi^2/$d.o.f. of order 1 is reached the 
results for $\eta$ do not match within statistical errors. This is 
a reminder that a small $\chi^2/$d.o.f. does not imply 
that systematical errors are of the same size as the statistical ones.

Since the statistical error with $\xi_{2nd}/L$ fixed is smaller we take 
our final result from these fits.  
In order to estimate systematical errors we compare  
results of
fits with the range $L_{min},L_{max}$ and
$L_{min}'=2 L_{min}$, $L_{max}'=2 L_{max}$. Then the error due to
$L^{-2}$ (which we assume to be the leading corrections beyond 
$L^{-\omega})$ corrections in the second interval should be $1/3$ of the
difference of the two results (up to a difference in the distribution
of the data with the interval). As our final estimate we take the 
fit result from $L_{min}=14$ and $L_{max}=48$. For comparison we fitted
with $L_{min}=7$ and $L_{max}=24$. For this interval we get 
$\eta=0.03800(13)$. Hence the systematical error  
from $L^{-2}$ corrections should be smaller than $0.00012$ (taking 
statistical errors into account).

\begin{table}[h]
\caption[bbb11]
{\label{chixi2} \sl Fits of the magnetic susceptibility at 
$\xi_{2nd}/L = 0.5927$
fixed with the
extended ansatz (\ref{chifit2}).
}
\vskip 0.2cm
\begin{center}
\begin{tabular}{|c|c|c|c|c|} \hline
$L_{min}$  & $c$ & $d$ & $\eta$ & $\chi^{2} \mbox{/d.o.f.}$ \\ \hline
\phantom{0}6  & -0.3602(40) & 1.26187(21) & 0.03784(5)\phantom{0}& 2.07 \\
\phantom{0}7  & -0.3809(68) & 1.26246(26) & 0.03798(6)\phantom{0}& 1.33 \\
\phantom{0}8  & -0.395(11)\phantom{0}& 1.26280(34)&0.03805(8)\phantom{0}
& 1.17 \\
10 & -0.381(18)\phantom{0}  & 1.26254(43) & 0.03800(10) & 1.06 \\
12 & -0.393(32)\phantom{0}  & 1.26275(62) & 0.03804(14) & 1.21 \\
14 & -0.405(43)\phantom{0}  & 1.26289(73) & 0.03807(16) & 1.40 \\
16 & -0.436(72)\phantom{0}  & 1.26330(99) & 0.03815(21) & 1.32 \\
\hline
\end{tabular}
\end{center}
\end{table}

\begin{table}
\caption[bb1r]
{\label{chiu2} \sl Fits of the magnetic susceptibility at
$U = 1.243$ 
fixed with the
extended ansatz (\ref{chifit2}).
}
\vskip 0.2cm
\begin{center}
\begin{tabular}
{|c|c|c|c|c|} \hline
$L_{min}$  & $c$ & $d$ & $\eta$ & $\chi^2/\mbox{d.o.f.}$ \\ \hline
\phantom{0}4  & -0.464(4)\phantom{0}& 1.2651(4)\phantom{0}& 
  0.03845(11) &  3.27 \\
\phantom{0}6 & -0.525(12) & 1.2681(6)\phantom{0}& 0.03917(16) &  0.76 \\
\phantom{0}8 & -0.526(30) & 1.2682(10) & 0.03918(23) &  0.85 \\
 10 & -0.553(55) & 1.2688(14) & 0.03931(31) &  0.96 \\
 12 & -0.574(90) & 1.2691(18) & 0.03937(40) &  1.05 \\
 14 & -0.47(13)\phantom{0}  & 1.2676(22) & 0.03907(49) &  1.17 \\
 16 & -0.24(23)\phantom{0}  & 1.2649(32) & 0.03851(67) &  1.27 \\
\hline
 \end{tabular}
\end{center}
\end{table}

Finally we checked for systematical errors due to residual leading order
corrections to scaling at $\lambda=2.0$.  For this purpose we fitted 
our data for $\lambda=1.0$ and $\lambda=4.0$ also with $L_{min}=14$ and 
ansatz (\ref{chifit2}).
We get $\eta=0.0375(13)$ and $\eta=0.0373(13)$ respectively. Taking into
account the statistical errors we find that 
\begin{equation}
\left| \frac{\Delta \eta_{eff}}{\Delta c_1(\lambda)} \right| < 0.018
\;\;\; .
\end{equation}

From the previous section we know that the coefficient 
$c_1(2.0)$ should be smaller than $0.007$. (Taking the fit result for 
$L_{min}=16$ plus the statistical error).
Therefore the systematical error in our
final estimate of $\eta$ due to residual leading order corrections should
be smaller than 0.00013. 
As a check we repeated the error-analysis along the 
lines of ref. \cite{us} and came up with a similar estimate.

As final estimate for $\eta$ we take the result from fitting the magnetic 
susceptibility  at $\xi_{2nd}/L$ fixed with the ansatz 
(\ref{chifit2}) and $L_{min}=14$
\begin{equation}
\eta=0.0381(2)[2] \;\;.
\end{equation}
The estimate of the systematical error is given in the second bracket. It
covers  residual $L^{-\omega}$ corrections and higher order corrections.

\subsection{The exponent $\nu$}
We computed
the derivate of the Binder cumulant $U$ with respect to $\kappa$ at the
fixed value of the Binder cumulant $U=1.243$ and at the fixed value
of $\xi_{2nd}/L=0.5927$. These quantities behave as 
\begin{equation}
\label{simplenu}
\overline{\frac{\partial U}{\partial \kappa}} =  c \;\; L^{1/\nu} \;\; .
\end{equation}
Results of the fits  are summarised in table \ref{abx} and \ref{abu}
for fixed $\xi_{2nd}/L$ and for fixed $U$, respectively. The 
$\chi^2/$d.o.f. becomes order 1 starting from $L_{min}=8$ and 
$L_{min}=7$ respectively. The statistical errors are slightly smaller
in the case of fixed $U$.

\begin{table}[h]
\caption[bbrx]
{\label{abx} \sl Fits of $\frac{\partial U}{\partial \kappa}$
at $\xi_{2nd}/L$ fixed
with the ansatz (\ref{simplenu}).
 }
\vskip 0.2cm
\begin{center}
\begin{tabular} {|c|c|c|c|} \hline
  $L_{min}$  & $c/2$ & $\nu$ & $\chi^2/\mbox{d.o.f.}$ \\ \hline
\phantom{0}6 & -0.5542(5)\phantom{0}  & 0.6709(1)  & 3.82 \\
\phantom{0}7 & -0.5565(6)\phantom{0}  & 0.6715(2)  & 1.76 \\
\phantom{0}8 & -0.5578(7)\phantom{0}  & 0.6719(2)  & 1.02 \\
  10 & -0.5586(9)\phantom{0}  & 0.6721(2)  & 0.96 \\
  12 & -0.5595(11) & 0.6723(3)  & 0.80 \\
  14 & -0.5608(13) & 0.6727(4)  & 0.65 \\
  16 & -0.5620(18) & 0.6729(5)  & 0.69 \\
  20 & -0.5610(24) & 0.6727(6)  & 0.63 \\
  24 & -0.5632(36) & 0.6732(9)  & 0.50 \\
 \hline
\end{tabular}
\end{center}
\end{table}

\begin{table}
\caption[brxz]
{\label{abu} \sl Fits of $\frac{\partial U}{\partial \kappa}$ 
at $U$ fixed with the 
ansatz (\ref{simplenu}).}
\vskip 0.2cm
\begin{center}
\begin{tabular} {|c|c|c|c|} \hline
$L_{min}$  & $c/2$ & $\nu$ & $\chi^2/\mbox{d.o.f.}$ \\ \hline
\phantom{0}6  &   -0.5551(4)\phantom{0}  &  0.6712(1)   &   2.07 \\
\phantom{0}7  &   -0.5564(5)\phantom{0}  &  0.6716(1)   &   1.19 \\
\phantom{0}8  &   -0.5572(6)\phantom{0}  &  0.6718(2)   &   0.80 \\
 10     &     -0.5577(7)\phantom{0}  &  0.6719(2)   &   0.73 \\
 12     &     -0.5583(9)\phantom{0}  &  0.6721(3)   &   0.62 \\
 14     &     -0.5593(11)  &    0.6723(3)   &   0.51 \\
 16     &     -0.5601(15)  &    0.6725(4)   &   0.55 \\
 20     &     -0.5592(21)  &    0.6723(5)   &   0.48 \\
 24     &     -0.5610(31)  &    0.6727(7)   &   0.30  \\
\hline
\end{tabular}
\end{center}
\end{table}

As in the case of the exponent $\eta$ we expect 
in addition to the statistical error  systematical errors 
due to the fact that the coefficient of $L^{-\omega}$ corrections 
does not vanish exactly and due to sub-leading $L^{-2}$ corrections.

In order to estimate these errors we proceed as in the previous section.

As our final result we take the fit with $L_{min}=14$ and $L_{max}=48$ 
of $\frac{\partial U}{\partial \kappa}$ at fixed $U$.
In order to estimate $L^{-2}$ corrections we
fitted  the data in the interval 
$L_{min}=7$ and $L_{max}=24$. For these lattices sizes we 
obtain $\nu=0.6712(2)$. Hence 
the estimate for a $L^{-2}$ error is $0.0011(5)/3 \approx 0.0005$.

In order to estimate the error due to residual 
$L^{-\omega}$ corrections we 
fitted our data for $\lambda=1.0$ and $\lambda=4.0$.  From $L_{min}=14$
we obtain $\nu=0.6706(11)$ for $\lambda=1.0$ and $\nu=0.6758(10)$ for 
$\lambda=4.0$. 
Hence 
\begin{equation}
\left| \frac{\Delta \nu_{eff}}{\Delta c_1(\lambda) } \right|
< 0.04 \;\;\; .
\end{equation}
From the previous section we know that 
$ c_1(2.0) \approx 0.007 \;$.
Therefore the estimate of the systematical error in $\nu$ is
$0.04 \times 0.007 \approx 0.0003 \;$ .

We arrive at our final estimate
\begin{equation}
\nu=0.6723(3)[8] \;\; ,
\end{equation}
where the statistical error is given in the first bracket and 
the systematical error that is given in the second bracket covers
$L^{-2}$ and residual $L^{-\omega}$ corrections.

\section{Comparison with the literature}
In table \ref{literature} we give for comparison recent results for 
critical exponents. 
Critical exponents for the XY-universality
class have been calculated using the high temperature series expansions, 
the $\epsilon$-expansion, perturbation theory in three dimension and 
Monte Carlo simulations. Our result for $\nu$ is 
consistent within error-bars
with (almost) all other 
theoretical results given in table \ref{literature}.
The result of the MC study
\cite{mc1} seems to be a little too small. Our error-bar is
smaller than that of all previous estimates.
Our estimate for
$\eta$ is consistent with the other theoretical estimates except with the 
Monte Carlo results. The values of refs. \cite{mc0,mc1} are too small
compared with our present estimate. Note that in these studies no 
careful check of systematical errors due to corrections to scaling was 
performed.
On the other hand the result of ref. \cite{mc2}, which takes into account
$L^{-\omega}$ corrections, is  by two standard 
deviations larger than our result.

In contrast to the one-component case \cite{MH} our result for the 
correction to scaling exponent $\omega$ is consistent with that obtained
with field theoretic methods.

\begin{table}[h]
\caption[bx10]
{
\label{literature} \sl
Recent results for critical exponents obtained with Monte Carlo
simulations (MC), $\epsilon$-expansion, Perturbation-Theory
in three dimensions
(3D,PT) and High temperature series expansions.   When only $\nu$ and 
$\gamma$ are given in the reference we computed $\eta$ with the scaling 
law. These cases are indicated by $^*$. In ref. \cite{helium1} a result
for $\alpha$ is given. In the table it is converted to $\nu$ using 
the scaling relation $\alpha=2-d \nu$. 
For a discussion see the text.
}
\vskip 0.2cm
\begin{center}
\begin{tabular}{|c|c|l|l|l|}
\hline
Ref.               & Method &\mc{1}{|c|}{$\nu$} & \mc{1}{|c|}{$\eta$} & 
		    \mc{1}{|c|}{$\omega$} \\
\hline
present work    & MC  & 0.6723(3)[8] & 0.0381(2)[2] & 0.79(2) \\
\cite{mc0}       & MC & 0.670(2)& $0.025(7)^*$ & \\ 
\cite{mc1}       & MC    & 0.662(7)   & 0.026(6)   &   \\
\cite{mc2}        & MC      & 0.6721(13) & 0.042(2) &  \\
\cite{guzi}      & 3D,PT    & 0.6703(15) & 0.0354(25) & 0.789(11)  \\
\cite{guzi}      & $\epsilon$,bc &0.6680(35) & 0.0380(50)&  0.802(18) \\
\cite{guzi}      & $\epsilon$,free &0.671 & 0.0370 & 0.802(18) \\
\cite{buco}      & HT & 0.674(2)  & $0.039(7)^*$ & 
  \\ 
\cite{helium1} & $^4$He & 0.67095(13) & 
&  \\ 
\cite{helium2} & $^4$He & 0.6705(6)  &           &   \\
\cite{helium3} & $^4$He & 0.6708(4)  &           &  \\
\hline
\end{tabular}
\end{center}
\end{table}

Experimental results for the exponent $\nu$ 
have been obtained for  the $\lambda$-transition of $^4$He. These results
have smaller error-bars than our Monte Carlo result. The experimental 
results are all smaller then our value but still the error-bars
touch.

\section{Conclusion and outlook}
In this paper we have improved the accuracy of the theoretical 
estimate of $\nu$ of the 3D XY universality class considerably.
In particular we give in addition to the statistical error 
a careful estimate of systematical errors that are caused by 
corrections to scaling.
Our value $\nu=0.6723(3)[8]$ is consistent with other theoretical 
estimates. However it is larger than the experimental results 
obtained from the $\lambda$-transition of $^4$He 
\cite{helium1,helium2,helium3} that give values from 
$0.6704$ up to $0.6709$
with an error in the last digit.  
It would be interesting to further improve the theoretical estimate 
to the claimed accuracy of the experimental results. This could be 
achieved by simulating at our best estimate for $\lambda_{opt}=2.1$ 
and going to linear lattice sizes roughly twice as large as in the 
present study to reduce the effect of sub-leading corrections.
At a sustained statistical accuracy this would require about 10 years
of CPU-time on a modern PC. 

In addition to critical exponents amplitude ratios are universal and 
have been experimentally determined for the $\lambda$-transition of $^4$He.
For example the specific heat behaves in the 
neighbourhood of the phase transition as 
\begin{equation}
C = A_{\pm} \; |t|^{-\alpha} \; (1 +  D_{\pm} \; |t|^{\theta} 
+ E_{\pm} \; t ) \;+\; B  \;\;,
\end{equation}
where $t=(T-T_c)/T_c$ 
is the reduced temperature.
The constants $A_{\pm}$, $D_{\pm}$, $E_{\pm}$ and $B$ depend on 
the system that is considered. The subscript $\pm$ indicates the low and 
high temperature phase. However renormalization group predicts 
the ratio
$A_+/A_-$ to be universal.  
Setting $\lambda=\lambda_{opt}$ leads to $D_{\pm}=0$ which greatly simplifies 
the determination of $A_+/A_-$ in a Monte Carlo simulation.

\end{document}